\documentclass[conference]{IEEEtran}
\ifCLASSINFOpdf
\else
\fi

\usepackage{amssymb}
\usepackage{setspace}
\usepackage{graphicx}
\usepackage[tight,footnotesize]{subfigure}
\usepackage{booktabs}
\usepackage{amsthm}
\usepackage{amsmath}
\usepackage{marvosym}
\usepackage{multirow}
\usepackage{slashbox}
\usepackage{cite}
\usepackage{array}

\def\BibTeX{{\rm B\kern-.05em{\sc i\kern-.025em b}\kern-.08em T\kern-.1667em\lower.7ex\hbox{E}\kern-.125emX}}

\allowdisplaybreaks[4]
\begin{document}

\title{A Traceable Concurrent Data Anonymous Transmission Scheme for Heterogeneous VANETs}

\author{\IEEEauthorblockN{Jingwei Liu\IEEEauthorrefmark{1},
Qin Hu\IEEEauthorrefmark{1},
Chaoya Li\IEEEauthorrefmark{1},
Rong Sun\IEEEauthorrefmark{1},
Xiaojiang Du\IEEEauthorrefmark{2}, and
Mohsen Guizani\IEEEauthorrefmark{3}
}
\IEEEauthorblockA{\IEEEauthorrefmark{1}State Key Lab of ISN, Xidian University, Xi'an, 710071, China.\\ Email: jwliu@mail.xidian.edu.cn, 1390309647@qq.com, 527690544@qq.com, rsun@mail.xidian.edu.cn}
\IEEEauthorblockA{\IEEEauthorrefmark{2}Department of Computer and Information Sciences, Temple University, Philadelphia, PA 19122, USA.\\ Email: dxj@ieee.org}
\IEEEauthorblockA{\IEEEauthorrefmark{3}Department of Electrical and Computer Engineering, University of ldaho, Mosocow, ldaho, USA.\\ Email: mguizani@ieee.org}
}
\maketitle

\begin{abstract}
Vehicular Ad Hoc Networks (VANETs) are attractive scenarios that can improve the traffic situation and provide convenient services for drivers and passengers via vehicle-to-vehicle (V2V) and vehicle-to-infrastructure (V2I) communication. However, there are still many security challenges in the traffic information transmission, especially in the intense traffic case. For ensuring the privacy of users and traceability of vehicles, we propose a traceable concurrent data anonymous transmission scheme for heterogeneous VANETs. The scheme is based on certificateless aggregate signcryption, so it supports batch verification. Moreover, conditional anonymity is also achieved due to the involving of the pseudo-ID technique. Furthermore, it is a pairing-free scheme for the merit of multi-trapdoor hash functions. As a result, the total computation overhead is greatly reduced.
\end{abstract}
\IEEEpeerreviewmaketitle

\section{Introduction}\label{s_1}
Nowadays, VANETs have become more and more widely used \cite{Zeadally2012Vehicular} in smart traffic. As a part of the intelligent transportation system, it can be integrated into IoT and play an important role to improve traffic conditions in smart city. The entities in VANETs can exchange information via Dedicated Short Range Communication (DSRC) or Transport Layer Security (TLS) protocols in three communication modes including Vehicle-to-Vehicle (V2V), Vehicle-to-Infrastructure (V2I) and hybrid mode.

\begin{figure}[tb]
\begin{center}
\includegraphics[width=9cm]{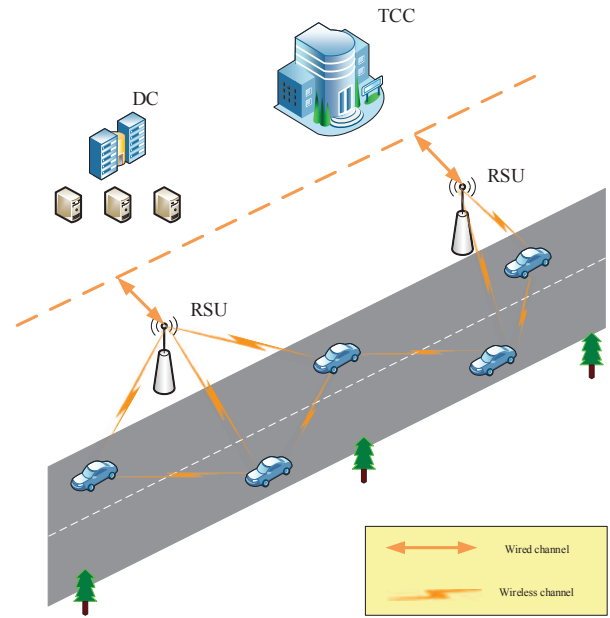}\\
\caption{A Scene of VANETs}
\end{center}
\end{figure}

A typical architecture of VANETs, as shown in Fig. 1, consists of On-Board Units (OBUs), Roadside Units (RSUs) and trusted authorities (TAs). OBUs embedded with sensors can identify static information (size, weight, etc. ) and collect dynamic information of vehicles (speed, direction, etc.). In addition, as both sources and routers, OBUs are able to send, receive and forward all kinds of information in VANETs  at any time. RSUs, a kind of infrastructures, are deployed at roadsides or crossroads. They can communicate with OBUs via wireless channel, as well as with other RSUs and TAs via wired channel. TAs are responsible for registering RSUs and OBUs, tracing vehicles involved in accidents and maintaining other services in VANETs. Aiming at the issue of the authentication of big data, there have been many batch authentication schemes specially designed for VANETs \cite{du2007effective,Zhang2008An,Du2008Security,Shim2012,Horng2015An,Zhang2015Toward,Lo2016An,Jiang2016An,Zhang2017Distributed}.

In 2003, the concept of aggregate signature was first introduced by D. Boneh et al. in \cite{boneh2003aggregate}. By taking the advantages of certificateless public key cryptosystem (CLC), many certificateless aggregate signature schemes were proposed \cite{xiong2013efficient,nie2016nclas}. An identity based aggregate signature scheme without pairing was then proposed for V2I communication \cite{Lo2016An}. In 1997, signcryption was first proposed by Y. Zheng in \cite{zheng1997digital}, which provided both public key encryption and digital signature in a single logic step. In 2010, Sun et al. proposed two heterogeneous signcryption schemes \cite{sun2010efficient} that achieved mutual secure communication between PKI and IBC. It was the first work on heterogeneous cryptosystem. In 2009, the first aggregate signcryption was proposed by Selvi et al. in \cite{selvi2009identity} that reduced computation overhead greatly. It allowed distinct signcryption cipertexts sent to the same recipient to be validated only once with the same security level. In recent years, many aggregate signcryptions have been raised \cite{babamir2012data,kar2013provably,Eslami2014Certificateless,han2014schap,basudan2017privacy}. However, the number of aggregate signcryption schemes is relatively small, in which some issues on security and efficiency still exist that remain to be solved. In 2015, Chandrasekhar et al. proposed an aggregate signcrytion scheme \cite{chandrasekhar2015efficient} based on a multi-trapdoor hash function \cite{chandrasekhar2014multi}.

In this paper, we mainly focus on two types of traffic data: unencrypted data and encrypted data. The unencrypted data is usually for vehicles to quickly obtain feedback from RSUs. The encrypted data is for the IoT data center to prevent opponents from eavesdropping or abusing these sensitive information. Unlike the first type of data, the second type of data often exists in the case of high traffic density. Therefore, based on an aggregate signcryption scheme with multi-trapdoor hash function, a secure data transmission scheme was proposed for V2I scenes. Our contributions are summarized as follows.

\begin{enumerate}
\item   Our scheme not only achieves batch verification of vehicles from OBUs to RSUs, but also accomplishes confidentiality and authentication in a single logic step.
\item   Based on multi-trapdoor hash functions, our scheme only involves scalar multiplications of fixed number without any bilinear pairing operations.
\item   The aggregate verification information could be validated without the plaintexts and the intended receiver. Therefore, it wouldn't take extra computation on decryption, once the batch verification is invalid.
\end{enumerate}

This paper is organized as follows. In section II, we briefly introduce some preliminaries. In section III, a traceable concurrent anonymous transmission scheme is constructed for VANETs. In section IV, we show the superiority of the proposed scheme by evaluating performance. Finally, the conclusion is given in section V.

\section{Preliminaries}
\label{s_2}

\subsection{Multi-trapdoor hash functions}
\label{s_2_2}
Let $\langle x, y\rangle_{[a,b]}$ denote a set of parameters $x_i$ and $y_i$($i\in[a,b]$), $\langle x \to {x^{'}},y \to {y^{'}}\rangle _{[a,b]|\{ i\}}$ denote $\langle x, y\rangle _{[a,b]}$, but the $i$th pair $\langle x_i, y_i\rangle$ is updated by a new one. The concept of multi-trapdoor hash function \cite{chandrasekhar2015efficient} is described as follows:

Assuming that there exist $n$ participants $\{ {u_i}\} _{i = 1}^n$, each of them have their own key pair $\langle T{K_i}, H{K_i}\rangle$. $HK$ is the public key (or hash key) for generating the multi-trapdoor hash value, and $TK$ is the private key (or trapdoor key) that is kept securely to generate the multi-trapdoor hash function collision by its owner. Concretely, each $u_i$ takes a message $m_i$, a random value $r_i$ and its public key $HK_i$ as the input, then calculates a multi-trapdoor hash value $T{H_{ \langle HK\rangle _{[1,n]}}}( \langle m,r\rangle _{[1,n]})$ as the output. In addition, with the corresponding trapdoor key $TK_i$, another chosen message $m_i'$ and a random value $r_i'$, each $u_i$ can construct an ephemeral hash key $HK_i'$ such that $T{H_{\langle HK\rangle _{[1,n]}}}( \langle m,r\rangle _{[1,n]}) = T{H_{\langle HK \to H{K'}\rangle _{[1,n]|\{ i\} }}}(\langle m \to {m'},r \to {r'}\rangle _{[1,n]|\{ i\} })$. When all of the participates generate their collision parameters respectively, a multi-trapdoor hash collision value can be calculated, namely $T{H_{\langle HK\rangle _{[1,n]}}}(\langle m,r\rangle _{[1,n]}) = T{H_{\langle HK \to H{K'}\rangle_{[1,n]|\{ 1,2,3, \cdot  \cdot  \cdot ,n\} }}}(\langle m \to {m'},r \to {r'}\rangle _{[1,n]|\{ 1,2,3, \cdot  \cdot  \cdot ,n\}})$.

\subsection{Security Assumptions}
\label{s_2_3}
Let $G$ be a set of points on an elliptic curve over a finite field $(E/F_q)$. The security of the proposed scheme is dependent on  the following security assumptions:

\noindent\textbf{Definition 1.} Elliptic Curve Discrete Logarithm Problem (ECDLP): Let $P$ be a point on an elliptic curve over a finite field. Given a random instance $(P,aP)$ for any $a \in Z_q^*$, it is difficult to compute $a$.

\noindent\textbf{Definition 2.} Computational Diffie-Hellman Problem (CDHP): Let $P$ be a generator of $G$. Given a random instance $\langle P,aP,bP\rangle$ for any $a,b \in Z_q^*$, it is difficult to compute $abP$.

\section{The Traceable Concurrent Data Anonymous Transmission Scheme for VANETs} \label{s_3}

\subsection{System Architecture} \label{s_2_1}
Fig. 2 shows the overview of the VANETs system architecture, including four entities: the OBUs equipped on the vehicles, the RSUs at the roadside, a trace authority (TRA) and a key management center (KMC).

The OBUs with constrained computing ability are responsible for transmitting the traffic-related data to the RSUs. The data should be signcrypted with the parameters stored in a tamper-proof device (TPD) that ensures the OBUs cannot be compromised. GPS equipped in OBUs can provide the precise localization. The RSUs with more computing power collect the sigcrypted traffic information from vehicles and unsigncrypt them in an aggregate manner. If the verification is valid, the RSUs will send feedback to vehicles and forward the traffic information to the traffic data center. The KMC is an honest but curious authority that is responsible for issuing certificates for the RSUs in PKI and generating partial private keys for vehicles in CLC. Different works \cite{Du2009A,xiao2007survey,Du2005Designing,Zhang2015Interference} have studied on the relevant security issues of key management. Furthermore, in order to achieve conditional anonymity in smart traffic, the TRA plays the role of a trusted authority who is in charge of generating the pseudo-ID and tracing malicious vehicles. Different from OBUs and RSUs that are online, the KMC and the TRA are offline in the registration stage and the trace stage respectively. According to the IEEE 802.11p standard, the OBUs and RSUs communicate with each other via a wireless communication protocol--DSRC, while the RSUs interact with the traffic data center via a wired protocol--TLS.

\subsection{Design Goals}

To meet the security demands in VANETs, the proposed scheme can provide the following properties:
\begin{itemize}
\item {\bfseries Confidentiality}. Before sent to the nearby RSUs, the traffic information from OBUs should be encrypted to keep the opponents from eavesdropping and analyzing the further attacks. So, we deploy the aggregate signcryption scheme as the cryptographic essential for confidentiality.
\item {\bfseries Conditional anonymity}. The identities of OBUs, such as the plate number, are often involved in the secure communication between OBUs and RSUs, which might lead to privacy issues. So it is necessary for OBUs to adopt the pseudo-IDs of participators instead of the real identities in the whole protocol.
\item {\bfseries Key escrow freeness}. Our scheme adopts the certificateless technique to manage vehicles so that they can generate their own secret keys, avoiding key escrow problem.
\item {\bfseries Low computational overhead}. In general, bilinear pairings are the most time-cost cryptographic operations in a security protocol. Based on an improved multi-trapdoor hash function, the proposed scheme achieves batch verification on the OBUs' report without pairing operations, so the performance of the new scheme is improved greatly.
\end{itemize}

\begin{figure}[tb]
\begin{center}
\includegraphics[width=9.5cm]{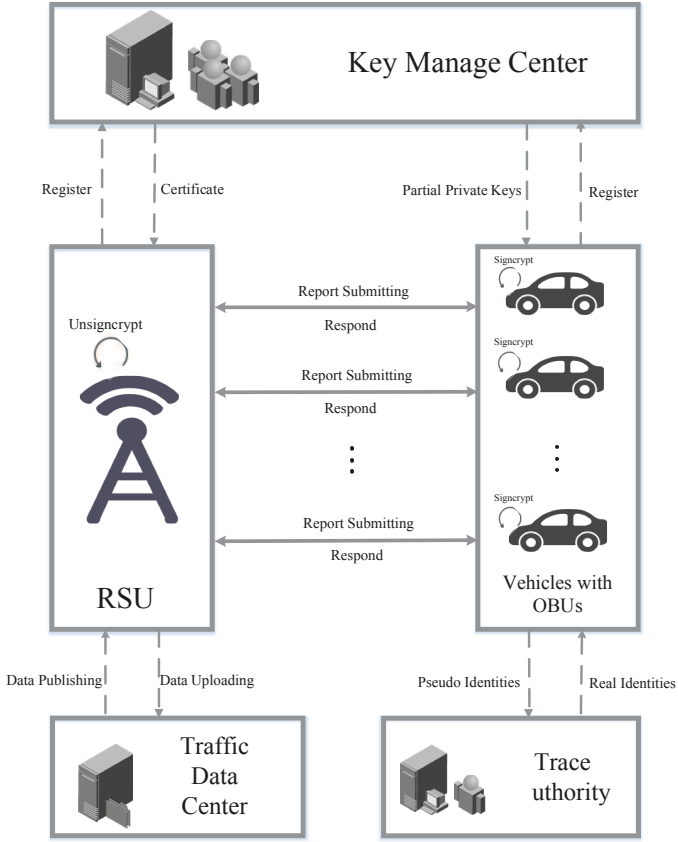}\\
\caption{A System Architecture for VANETs} \label{architecture}
\end{center}
\end{figure}

\subsection{The Protocol}

In this section, the traceable concurrent anonymous transmission scheme for heterogeneous VANETs will be introduced in detail. It consists of the following seven phases: system initialization, pseudo-ID generation, vehicle registration, RSU registration, RSU broadcast, traffic information uploading, batch verification and decryption. The detailed processes are described as follows.

\subsubsection{System Initialization}

\begin{itemize}
\item	Both $KMC$ and $TRA$ choose a same elliptic curve $E$ over a finite field $F_q$. Let $G$ be an additive group that consists of the points on $E/F_q$, $P$ be a generator of $G$, $O$ denote infinity, where $P \ne O$.
\item	$KMC$ randomly chooses $s \in Z_q^*$ as its master secret key and computes ${P_{pub}} = sP$ as the master public key.
\item   $TRA$ randomly chooses $\beta \in Z_q^*$ as its secret key and computes ${T_{pub}} = \beta P$ as its public key.
\item   Both $KMC$ and $TRA$ choose two one-way cryptographic hash functions: $H_1: {\{ 0,1\} ^*} \to Z_q^*$, ${H_2}:{\{ 0,1\} ^*} \to {\{ 0,1\} ^{{l_m}}}$, where $l_m$ denotes the  bit length of the messages.
\item   $KMC$ publishes the system parameter $params = \langle q,k,E/F_q,G,P,P_{pub},T_{pub},H_1,H_2\rangle$. Note that $params$ will be stored by $V_i$ and RSUs before their registration.
\end{itemize}

\subsubsection{Pseudo-ID Generation}
\begin{itemize}
\item	$V_i$ randomly chooses ${\lambda _i} \in Z_q^*$ and computes $PID_{i1}={\lambda _i}P$ as one part of the pseudo-ID of $V_i$.
\item	$V_i$ sends $\langle RID_i, PID_{i1}\rangle$ to $TRA$ via a secure channel, where $RID_i$ is the real identity of $V_i$.
\item	$TRA$ computes \[PI{D_{i2}} = RI{D_i} \oplus H_1(\beta PI{D_{i1}},PI{D_{i1}},T_i),\] where $T_i$ is the period of validity of the pseudo-ID. Then it sends $PI{D_i} = \langle PI{D_{i1}},PI{D_{i2}},T_i\rangle$ to $V_i$ via a secure channel, where $PI{D_{i2}}$ is the other part of the pseudo-ID of $V_i$.
\item   $V_i$ sets $PID_i$ as its pseudo-ID.
\end{itemize}

\begin{table*}[tp]
\centering
\caption{Complexity Comparison}
\begin{tabular}{ccccccc}  
\toprule
The scheme                                          & Cryptosystem  & Application & Signcryption & Verification      & Decryption \\   
\midrule
J. Kar \cite{kar2013provably}                       & IBC           & Theory      & 1P+2M+3H     & 2H+4nM            & nP+nH      \\ \relax         
Ziba Eslami et al. \cite{Eslami2014Certificateless} & CLC           & Theory      & 1P+3M+2MH+1H & (n+3)P+(n+1)MH    & nP+nM+nH   \\ \relax
Y. Han et al. \cite{han2014schap}                   & PKI           & VANETs      & 3M+1MH+1H    & (n+1)P+nMH        & nM+nH      \\ \relax        
S. Basudan et al. \cite{basudan2017privacy}         & CLC           & VANETs      & 7M+3H        & (n+4)P+nM+(n+1)H  & 2nM+nH     \\
Ours                                                & Heterogeneous & VANETs      & 3M+3H        & 5M+(3n+1)H        & nM+nH      \\
\bottomrule
\end{tabular}
\end{table*}

\subsubsection{Vehicle registration}
\begin{itemize}
\item	$V_i$ randomly chooses $l_i \in Z_q^*$ as its secret value and computes $L_i=l_iP$.
\item	$KMC$ randomly chooses $k_i \in Z_q^*$, computes $Y_i=k_iP$ as $V_i$'s partial public key, where $Q_i = H_1(PID_i,Y_i)$, and computes ${y_i} = {k_i} + sQ_i$ as $V_i$'s partial private key. Finally, $KMC$ sends $y_i$ and $Y_i$ to $V_i$ via a secure channel.
\item	$V_i$ sets $\langle y_i,l_i\rangle$ as its full private key, $\langle Y_i,L_i\rangle$ as its full public key.
\end{itemize}

\subsubsection{RSU registration}
\begin{itemize}
\item	$R$ randomly chooses ${\gamma } \in Z_q^*$  and computes its public key $Y_R={\gamma} P$. Then, $R$ sends its identity $ID_R$ and $Y_R$ to $KMC$ via a secure channel.
\item	$KMC$ generates a certificate ${cert_R} = \langle ID_R, Y_R, Sig_R\rangle$ for $R$, where $Sig_R$ is a signature signed by $KMC$'s master secret key.
\item   $KMC$ sends $cert_R$ to $R$ via a secure channel. 	
\end{itemize}

\subsubsection{RSU broadcast}
$R$ chooses a public random number $\eta $ and generates a digital signature ${Sig\left( \eta,\gamma ,tt_R \right)}$ with its private key $\gamma$, where $tt_R$ denotes the timestamp. Then, $R$ constructs a packet $PKT = \left\langle {cert_R,\eta,Sig\left( {\eta ,\gamma ,tt_R} \right)} \right\rangle $ and broadcasts it periodically.

\subsubsection{Traffic Information Uploading}
\begin{itemize}
\item	When $V_i$ enters into the communication zone of $R$, it firstly checks the $cert_R$ broadcasted by $R$. If illegal, it aborts. Otherwise, $V_i$ continues to verify if the signature in the broadcast packet is valid. If not, it aborts. Otherwise, $V_i$ extracts $ID_R$ and $Y_R$ from $cert_R$ and goes to the next step.
\item	Choose $x_i \in Z_q^*$ randomly and compute ${X_i} = {x_i}P$.
\item	Collect traffic information $m_i$ and compute ${c_i} = {m_i} \oplus {H_2}({x_i}{Y_R})$.
\item   Compute the ephemeral trapdoor key ${z_i} = $\\$\eta({H_1}(PID_i,Y_i,L_i,tt_i) -  {H_1}({c_i},{X_i},{cert}_R,{tt}_i)) + ({y_i} + {l_i})$, where $tt_i$ denotes the timestamp.
\item   Compute $t_i=x_i-{H_1(\eta Y_R)}(z_i+l_i)$.
\item   Send the signcrypted information $${\sigma _i} = \left\langle {{t_i},{c_i},{X_i},{Y_i},{L_i},{PID_i},t{t_i}} \right\rangle$$ to $R$.
\end{itemize}

\subsubsection{Batch verification and decryption}

On receiving signcrypted messages $\sigma_i$ from $V_i$, $R$ does as follows:

\begin{itemize}
\item   Compute ${Q_i} = {H_1}\left( {PID_i},{Y_i} \right)$, $h_{3i}=H_1(PID_i,Y_i,L_i,tt_i)$ and ${h_{4i}} = {H_1}\left( {{c_i},{X_i},cer{t_R},t{t_i}} \right)$ for each $V_i$.
\item   Compute $t = \sum\limits_{i = 1}^n {{t_i}} $, $X = \sum\limits_{i = 1}^n {{X_i}} $, $L = \sum\limits_{i = 1}^n {{L_i}} $, where $n$ denotes the number of vehicles.
\item   Compute the ephemeral hash key $$Z = H_1^{-1}(y_R\eta P)\left( {X - tP} \right) - L.$$
\item   Compute the multi-trapdoor hash function value $$V_1 = \eta\left( {\sum\limits_{i = 1}^n {h_{3i}} } \right)P +  \sum\limits_{i = 1}^n {\left( {{Y_i} + {L_i}} \right)}.$$
\item   Compute the multi-trapdoor collision value $$V_2 = \eta\left( {\sum\limits_{i = 1}^n {{h_{4i}}} } \right)P + Z - \left( { \sum\limits_{i = 1}^n {{Q_i}} } \right){P_{pub}}.$$
\item   Check if the equation $V_1=V_2$ holds. If not, it aborts. Otherwise, it goes to the next step.
\item   Calculate the message ${m_i} = {c_i} \oplus {H_2}\left( {{y_R}{X_i}} \right)$ of $V_i$.
\end{itemize}

\subsection{Correctness}
We can validate the correctness of the proposed scheme through formula derivation below.

\noindent${V_2} = \eta \left( {\sum\limits_{i = 1}^n {{h_{4i}}} } \right)P +Z - \left( {\sum\limits_{i = 1}^n {{Q_i}} } \right){P_{pub}}$\\
$  = \eta \left( {\sum\limits_{i = 1}^n {{h_{4i}}} } \right)P + H_1^{-1}(y_R\eta P)\left( {X - tP} \right)\\ - L - \left( {\sum\limits_{i = 1}^n {{Q_i}} } \right){P_{pub}}$\\
$  = \eta \left( {\sum\limits_{i = 1}^n {{h_{4i}}} } \right)P + H_1^{-1}(y_R\eta P)\left( {\sum\limits_{i = 1}^n {\left( {{x_i} - {t_i}} \right)} } \right)P \\- \sum\limits_{i = 1}^n {{L_i}}  - \left( {\sum\limits_{i = 1}^n {{Q_i}} } \right){P_{pub}}$\\
      $= \eta \left( {\sum\limits_{i = 1}^n {{h_{4i}}} } \right)P + \left( {\sum\limits_{i = 1}^n {\left( {{z_i} + {l_i}} \right)} } \right)P- \sum\limits_{i = 1}^n {{L_i}} \\ -
      \left( {\sum\limits_{i = 1}^n {{Q_i}} } \right){P_{pub}}$\\
      $= \eta \left( {\sum\limits_{i = 1}^n {{h_{4i}}} } \right)P + \left( {\sum\limits_{i = 1}^n {\left( {\eta \left( {{h_{3i}} - {h_{4i}}} \right) + {y_i} + {l_i}} \right)} } \right)P \\- \left( {\sum\limits_{i = 1}^n {{Q_i}} } \right){P_{pub}}$\\
      $= \left( {\eta \sum\limits_{i = 1}^n {{h_{3i}}}  + \sum\limits_{i = 1}^n {\left( {{k_i} + s{Q_i}} \right)}  + \sum\limits_{i = 1}^n {{l_i}} } \right)P - \left( {\sum\limits_{i = 1}^n {{Q_i}} } \right)sP$\\
      $= \left( {\eta \sum\limits_{i = 1}^n {{h_{3i}}}  + \sum\limits_{i = 1}^n {\left( {{k_i} + {l_i}} \right)} } \right)P\\$
      $= \eta \left( {\sum\limits_{i = 1}^n {{h_{3i}}} } \right)P + \sum\limits_{i = 1}^n {\left( {{Y_i} + {L_i}} \right)}\\
       = V_1$

\subsection{Security analysis}

In this section, we will discuss the security properties of the proposed scheme.

\subsubsection{Message authentication}

In the signcrytion stage, only the vehicle with the corresponding trapdoor key can generate an ephemeral trapdoor key, so the adversary cannot forge a valid signcryption unless he can solve the ECDLP. Furthermore, if the adversary attempts to recover the plaintexts from an aggregate signcryption, he has to encounter the CDHP obviously. Hence, our scheme achieves confidentiality, authentication, integrity and non-repudiation simultaneously.

\subsubsection{Internal security}

 None of OBU can impersonate any other OBUs to forge signcrypted messages, while a RSU can decrypt the signcrypted messages that are sent to other RSUs. Furthermore, even the $TA$ cannot forge a valid message of other entities in VANETs yet, since the certificateless cryptosystem is adopted in the registration stage. Hence, internal security is ensured in the proposed scheme.

\subsubsection{Conditional anonymity}

Because the pseudo-ID $PID_i$ is deployed for each $V_i$, the adversary cannot obtain any information about the actual identity $RID_i$ of $V_i$ without the secret key $\beta$ of $TRA$ that is used to generate the pseudo-ID during the data transmission process. If the adversary still attempts to reveal $V_i$'s real identity, it has to encounter the ECDLP that is assumed to be intractable.

\subsubsection{Traceability}

When a dispute on $V_i$ happens, only $TRA$ can extract the information of the real identity $RID_i$ of $V_i$ by calculating $$RI{D_i} = PI{D_{i2}} \oplus H_1\left( {\beta PI{D_{i1}},PI{D_{i1}},t{t_i}} \right),$$ which can make the traceability available.

\subsubsection{Unlinkability}

We claim that a secure protocol possesses unlinkability when there is no adversary that can judge if two different messages are from the same vehicle. Obviously, the proposed scheme can cover distinct messages by diverse pseudo-IDs and the corresponding private keys, because vehicles will update pseudo-IDs over a period of time. Therefore, the adversary cannot link different messages at different times to a specific vehicle, so that our scheme achieves unlinkability to a certain extent.  

\begin{figure}[tb]
\begin{center}
\includegraphics[width=8.5cm]{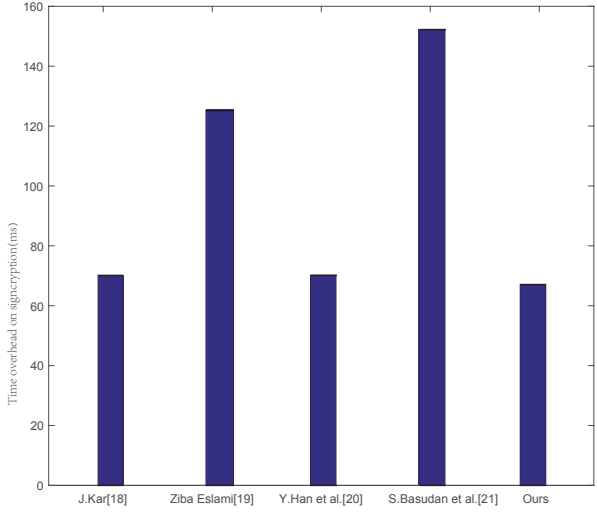}\\
\caption{Time Consumption on Signcryption} \label{signcryption}
\end{center}
\end{figure}

\begin{figure}[tb]
\begin{center}
\includegraphics[width=9cm]{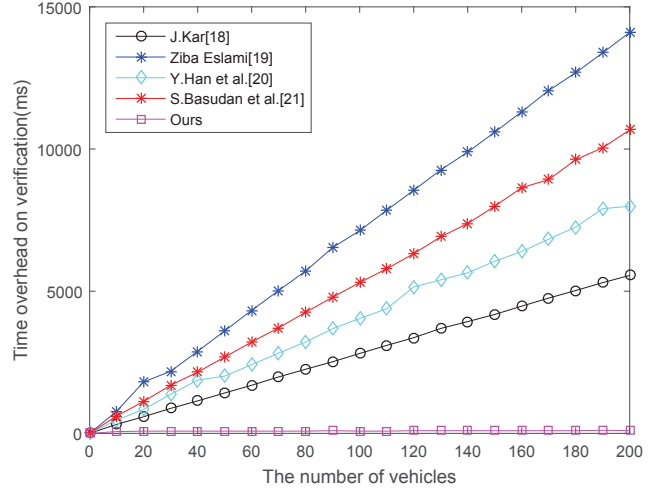}\\
\caption{Time Consumption on Verification} \label{verification}
\end{center}
\end{figure}

\begin{figure}[tb]
\begin{center}
\includegraphics[width=9cm]{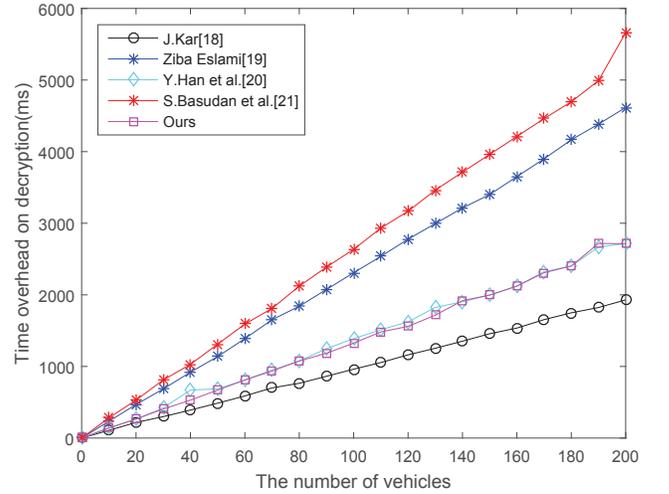}\\
\caption{Time Consumption on Decryption} \label{decryption}
\end{center}
\end{figure}

\begin{figure}[tb]
\begin{center}
\includegraphics[width=9cm]{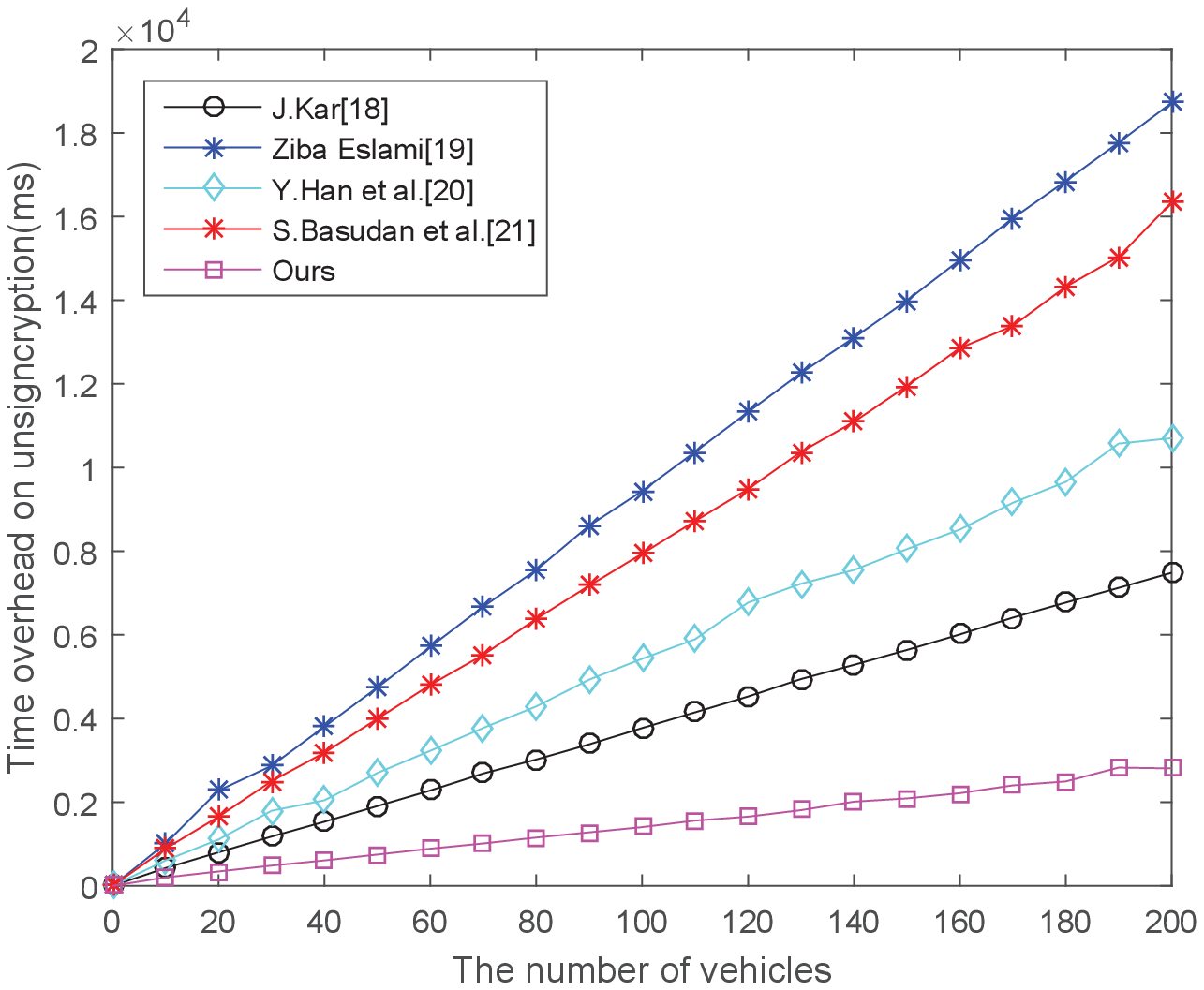}\\
\caption{Time Consumption on Unsigncryption} \label{unsigncryption}
\end{center}
\end{figure}

\section{Performance Evaluation}

In this section, we compare the proposed scheme with other existing relevant schemes in terms of the computation overhead. Firstly, for theoretical analysis of computation complexity, let $P$ denote the pairing operation, $M$ denote the scalar multiplication operation in $G_1$, $MH$ denote the MapToPoint hash operation, $H$ denote the general hash operation. Note that other mathematical operations such as additive operations in $G_1$ are omitted here since their influence is tiny in the performance evaluation.

As shown in Table I, the proposed scheme does not involve any pairing operations in all stages and only has five scalar multiplication operations in the verification stage that are independent on the number of vehicles, so it achieves the least cryptographic complexity on the whole compared with the other  four existing schemes, although the number of scalar multiplication operations in signcryption stage of ours is 2 more than that in Y. Han et al. \cite{han2014schap}.

In addition, in order to quantitatively analyze the computational efficiency of the proposed scheme, the simulation tests are performed over the type A elliptic curve in Java Pairing Based Cryptography 2.0.0 with an Intel G640 2.80 GHz processor. According to the results of the execution, the computational efficiency could be further analyzed. Fig. \ref{signcryption} demonstrates that the proposed scheme has a comparative advantage over other aggregate signcryption schemes on time-consumption in the signcryption stage. In Fig. \ref{verification}, the growing trend of the computation overhead in our scheme is the lowest in contrast to other schemes in the verification stage, because only low-complexity hash operations are associated with the number of the vehicles in our scheme. Although, in Fig. \ref{decryption}, the time overhead on decryption is a little more than that in J. Kar \cite{kar2013provably}, our scheme still achieves better performance than other schemes at the whole unsigncryption stage, as shown in Fig. \ref{unsigncryption}.The results show that our scheme is much more practical in VANETs application scenarios.

\section{Conclusion} \label{s_7}
In this paper, based on an improved aggregate signcryption scheme with multi-trapdoor hash functions, a traceable concurrent anonymous transmission scheme for heterogeneous VANETs was constructed. The confidentiality, integrity, authentication and non-repudiation are all achieved in a single logic step due to the merits of the aggregate signcryption algorithm. A pseudonym authority guaranteed the conditional anonymity of vehicles. Because of CLC cryptosystem is involved in the registration of vehicles, the heavy burden of certificate management of the KMC center is lightened. Most of all, it greatly decreases the computational overhead in batch verification stage by deploying multi-trapdoor hash functions instead of bilinear pairings. The proposed scheme vastly improves the flexibility and practicability of VANETs.

\section*{Acknowledgements}
This work is supported by the Key Program of NSFC-Tongyong Union Foundation under Grant U1636209, the 111 Project (B08038) and Collaborative Innovation Center of Information Sensing and Understanding at Xidian University.

\ifCLASSOPTIONcaptionsoff
  \newpage
\fi

\bibliographystyle{IEEEtran}
\bibliography{ms}

\begin{thebibliography}{10}
\providecommand{\url}[1]{#1}
\csname url@samestyle\endcsname
\providecommand{\newblock}{\relax}
\providecommand{\bibinfo}[2]{#2}
\providecommand{\BIBentrySTDinterwordspacing}{\spaceskip=0pt\relax}
\providecommand{\BIBentryALTinterwordstretchfactor}{4}
\providecommand{\BIBentryALTinterwordspacing}{\spaceskip=\fontdimen2\font plus
\BIBentryALTinterwordstretchfactor\fontdimen3\font minus
  \fontdimen4\font\relax}
\providecommand{\BIBforeignlanguage}[2]{{%
\expandafter\ifx\csname l@#1\endcsname\relax
\typeout{** WARNING: IEEEtran.bst: No hyphenation pattern has been}%
\typeout{** loaded for the language `#1'. Using the pattern for}%
\typeout{** the default language instead.}%
\else
\language=\csname l@#1\endcsname
\fi
#2}}
\providecommand{\BIBdecl}{\relax}
\BIBdecl

\bibitem{Zeadally2012Vehicular}
S.~Zeadally, R.~Hunt, Y.~S. Chen, A.~Irwin, and A.~Hassan, ``Vehicular ad hoc
  networks (vanets): status, results, and challenges,'' \emph{Telecommunication
  Systems}, vol.~50, no.~4, pp. 217--241, 2012.

\bibitem{du2007effective}
X.~Du, Y.~Xiao, M.~Guizani, and H.-H. Chen, ``An effective key management
  scheme for heterogeneous sensor networks,'' \emph{Ad Hoc Networks}, vol.~5,
  no.~1, pp. 24--34, 2007.

\bibitem{Zhang2008An}
C.~Zhang, R.~Lu, X.~Lin, P.~H. Ho, and X.~Shen, ``An efficient identity-based
  batch verification scheme for vehicular sensor networks,'' in \emph{Proc. of
  IEEE INFOCOM'08}, 2008, pp. 246--250.

\bibitem{Du2008Security}
X.~Du and H.~H. Chen, ``Security in wireless sensor networks,'' \emph{IEEE
  Wireless Communications Magazine}, vol.~15, no.~4, pp. 60--66, 2008.

\bibitem{Shim2012}
K.~A. Shim, ``\protect{CPAS}: An efficient conditional privacy-preserving
  authentication scheme for vehicular sensor networks,'' \emph{IEEE
  Transactions on Vehicular Technology}, vol.~61, no.~4, pp. 1874--1883, 2012.

\bibitem{Horng2015An}
S.~J. Horng, S.~F. Tzeng, P.~H. Huang, X.~Wang, T.~Li, and M.~K. Khan, ``An
  efficient certificateless aggregate signature with conditional
  privacy-preserving for vehicular sensor networks,'' \emph{Information
  Sciences}, vol. 317, no.~C, pp. 48--66, 2015.

\bibitem{Zhang2015Toward}
H.~Zhang, Q.~Zhang, and X.~Du, ``Toward vehicle-assisted cloud computing for
  smartphones,'' \emph{IEEE Transactions on Vehicular Technology}, vol.~64,
  no.~12, pp. 5610--5618, 2015.

\bibitem{Lo2016An}
N.~W. Lo and J.~L. Tsai, ``An efficient conditional privacy-preserving
  authentication scheme for vehicular sensor networks without pairings,''
  \emph{IEEE Transactions on Intelligent Transportation Systems}, vol.~17,
  no.~5, pp. 1319--1328, 2016.

\bibitem{Jiang2016An}
S.~Jiang, X.~Zhu, and L.~Wang, ``An efficient anonymous batch authentication
  scheme based on hmac for vanets,'' \emph{IEEE Transactions on Intelligent
  Transportation Systems}, vol.~17, no.~8, pp. 2193--2204, 2016.

\bibitem{Zhang2017Distributed}
L.~Zhang, Q.~Wu, J.~Domingo-Ferrer, B.~Qin, and C.~Hu, ``Distributed aggregate
  privacy-preserving authentication in vanets,'' \emph{IEEE Transactions on
  Intelligent Transportation Systems}, vol.~18, no.~3, pp. 516--526, 2017.

\bibitem{boneh2003aggregate}
D.~Boneh, C.~Gentry, B.~Lynn, and H.~Shacham, ``Aggregate and verifiably
  encrypted signatures from bilinear maps,'' in \emph{Proc. of Eurocrypt'03},
  2003, pp. 416--432.

\bibitem{xiong2013efficient}
H.~Xiong, Z.~Guan, Z.~Chen, and F.~Li, ``An efficient certificateless aggregate
  signature with constant pairing computations,'' \emph{Information Sciences},
  vol. 219, pp. 225--235, 2013.

\bibitem{nie2016nclas}
H.~Nie, Y.~Li, W.~Chen, and Y.~Ding, ``Nclas: a novel and efficient
  certificateless aggregate signature scheme,'' \emph{Security and
  Communication Networks}, vol.~9, no.~16, pp. 3141--3151, 2016.

\bibitem{zheng1997digital}
Y.~Zheng, ``Digital signcryption or how to achieve cost (signature \&
  encryption) \textless\textless cost (signature)+ cost (encryption),'' in
  \emph{Proc. of Annual International Cryptology Conference}, 1997, pp.
  165--179.

\bibitem{sun2010efficient}
Y.~Sun and H.~Li, ``Efficient signcryption between tpkc and idpkc and its
  multi-receiver construction,'' \emph{Science China Information Sciences},
  vol.~53, no.~3, pp. 557--566, 2010.

\bibitem{selvi2009identity}
S.~S.~D. Selvi, S.~S. Vivek, J.~Shriram, S.~Kalaivani, and C.~P. Rangan,
  ``Identity based aggregate signcryption schemes,'' in \emph{Proc. of
  International Conference on Cryptology in India}, 2009, pp. 378--397.

\bibitem{babamir2012data}
F.~S. Babamir and Z.~Eslami, ``Data security in unattended wireless sensor
  networks through aggregate signcryption,'' \emph{KSII Transactions on
  Internet And Information Systems}, vol.~6, no.~11, pp. 2940--2955, 2012.

\bibitem{kar2013provably}
J.~Kar, ``Provably secure identity-based aggregate signcryption scheme in
  random oracles,'' \emph{IACR Cryptology ePrint Archive}, vol. 2013, p.~37,
  2013.

\bibitem{Eslami2014Certificateless}
Z.~Eslami and N.~Pakniat, ``Certificateless aggregate signcryption: Security
  model and a concrete construction secure in the random oracle model,''
  \emph{Journal of King Saud University - Computer and Information Sciences},
  vol.~26, no.~3, pp. 276--286, 2014.

\bibitem{han2014schap}
Y.~Han, D.~Fang, Z.~Yue, and J.~Zhang, ``Schap: The aggregate signcryption
  based hybrid authentication protocol for vanet,'' in \emph{Proc. of
  International Conference on Internet of Vehicles}, 2014, pp. 218--226.

\bibitem{basudan2017privacy}
S.~Basudan, X.~Lin, and K.~Sankaranarayanan, ``A privacy-preserving vehicular
  crowdsensing based road surface condition monitoring system using fog
  computing,'' \emph{IEEE Internet of Things Journal}, vol.~4, no.~3, pp.
  772--782, 2017.

\bibitem{chandrasekhar2015efficient}
S.~Chandrasekhar and M.~Singhal, ``Efficient and scalable aggregate
  signcryption scheme based on multi-trapdoor hash functions,'' in \emph{Proc.
  of 2015 IEEE Conference on Communications and Network Security (CNS)}, 2015,
  pp. 610--618.

\bibitem{chandrasekhar2014multi}
S.~Chandrasekhar, ``Multi-trapdoor hash functions and their applications in
  network security,'' in \emph{Proc. of 2014 IEEE Conference on Communications
  and Network Security}, 2014, pp. 463--471.

\bibitem{Du2009A}
X.~Du, M.~Guizani, Y.~Xiao, and H.~H. Chen, ``A routing-driven elliptic curve
  cryptography based key management scheme for heterogeneous sensor networks,''
  \emph{International Journal of Computer Technology \& Applications}, vol.~8,
  no.~3, pp. 1223--1229, 2009.

\bibitem{xiao2007survey}
Y.~Xiao, V.~K. Rayi, B.~Sun, X.~Du, F.~Hu, and M.~Galloway, ``A survey of key
  management schemes in wireless sensor networks,'' \emph{Computer
  Communications}, vol.~30, no. 11-12, pp. 2314--2341, 2007.

\bibitem{Du2005Designing}
X.~Du and F.~Lin, ``Designing efficient routing protocol for heterogeneous
  sensor networks,'' in \emph{Proc. of 24th IEEE International Performance,
  Computing, and Communications Conference}, 2005, pp. 51--58.

\bibitem{Zhang2015Interference}
H.~Zhang, S.~Chen, X.~Li, H.~Ji, and X.~Du, ``Interference management for
  heterogeneous networks with spectral efficiency improvement,'' \emph{IEEE
  Wireless Communications}, vol.~22, no.~2, pp. 101--107, 2015.

\end{thebibliography}

\end{document}